\documentclass[twocolumn,a4paper,10pt,showpacs,prl,aps,superscriptaddress,amsmath,amssymb,amsfonts,floatfix]{revtex4-1}

\usepackage{graphicx}
\usepackage{epsfig}
\usepackage{latexsym}

\begin{document}

% The following information is for internal review, please remove them for submission
%\leftline{INTERNAL DOCUMENT -- NOT FOR PUBLIC DISTRIBUTION}
%\leftline{Version ii as of \today}

% the following line is for submission, including submission to the arXiv!!
%\hspace{5.2in} \mbox{Fermilab-Pub-04/xxx-E}

\title{Scaling behavior and strain dependence of in-plane elastic properties of graphene}
\author{J. H. Los, A. Fasolino, and M. I. Katsnelson}
%\author{J. H. Los, A. Fasolino, and M. I. Katsnelson}
\affiliation{Radboud University, Institute for Molecules and Materials,
                Heyendaalseweg 135, 6525AJ Nijmegen, The Netherlands}
\date{\today}

\begin{abstract}
We show by atomistic simulations that, in the thermodynamic limit, the in-plane
elastic moduli of graphene at finite temperature vanish with system size $ L $ 
as a power law $ ~ L^{-\eta_u} $ with $ \eta_u \simeq 0.325 $, in agreement with 
the membrane theory. Our simulations clearly reveal the size and strain
dependence of graphene's elastic moduli, allowing comparison to experimental data.
Although the recently measured difference of a factor 2 between the asymptotic value
of the Young modulus for tensilely strained systems
and the value from {\it ab initio} calculations remains unsolved,
our results do explain the experimentally observed increase of more than a factor 2 
for a  tensile strain
of only a few permille. We also discuss the scaling of the Poisson ratio, for 
which our simulations disagree with the predictions of the 
self-consistent screening approximation.
\end{abstract}

\pacs{}
\maketitle
Mechanical and structural properties of graphene form an intriguing
and highly non-trivial aspect of its physics. The structure of
a two-dimensional (2D) material embedded in a 3D space gives
room to special features, related to large out-of-plane deformations,
in particular thermal ripples \cite{Meyer,Fasolino1,Abedpour,Los1,Thompson,Zan,Gao,review,book}
and static ripples and wrinkles \cite{Zhu}.
A crucial difference with  3D (or strictly 2D) crystals
is that, for graphene, out-of-plane atomic displacements
$h$ and in-plane displacements ${\bf u}$ have different wavevector dependence
of the energy cost in the long wavelength
limit $q \rightarrow 0$, namely $ \propto q^2 $ and $ \propto q $ respectively, the latter being
the normal behavior for acoustic phonons. Hence, at finite temperature,
the long wavelength out-of-plane fluctuations are much larger than
the in-plane ones, so that at some small wavevector $q$ the first anharmonic coupling term 
of the form $ u h^2 $ will dominate over
the ``normal'' harmonic terms $ u^2 $, with important consequences
for the elastic behavior\cite{review,book,Nelson1,Nelson2,Doussal,Gazit}.
In particular, the wavevector
dependence of the anharmonic coupling strength leads to expect a power law behavior of the size dependence of the
elastic properties.

Contrary to the temperature dependence \cite{Zakharchenko,Chen},
so far the size dependence of the in-plane elastic moduli
of graphene has been hardly studied nor measured \cite{Lee} until recent
experiments seem to indicate that such a size dependence does exist for
graphene \cite{Polin1,Polin2}.
From indentation experiments on graphene drums with
sizes of the order of 1 $ \mu m $,
the Young modulus $Y$
was found to vary between 250 N/m and 700 N/m with increasing strain \cite{Polin2},
the latter value being much higher than the currently accepted value for flat sheets of  
$\sim$ 340 N/m obtained in previous measurements \cite{Lee} or {\it ab initio} calculations
\cite{Sanchez}.

Here we study the size dependence of the in-plane elastic moduli
of graphene at room temperature $T$=300 K by means of atomistic Monte Carlo (MC)
simulations based on the realistic interatomic potential LCBOPII
\cite{Los2}, as used in previous works \cite{Fasolino1,Los1,Zakharchenko}.
We obtain explicit expressions
for the size and strain dependence of graphene's in-plane elastic moduli,
providing a benchmark and tools for the analysis of experiments
for systems of any size.

Theoretically, the mentioned size dependence has been
studied within the continuum elastic theory of thin plates and membranes,
described by the Hamiltonian \cite{Landau}:
\begin{equation}
\label{Ham1}
H = \frac{1}{2}\int \! d {\bf r} \! \left( \kappa (\nabla^2 h)^2
+ \lambda u_{\alpha\alpha}^2 + 2 \mu u_{\alpha\beta}^2 \right)
\end{equation}
where $ {\bf r} $ is the 2D position vector, $\kappa$ is the bending
rigidity, $\lambda$ and $\mu$ are Lam\'{e} coefficients, with $ \mu $
the shear modulus, and 
\begin{equation}
\label{strtens}
u_{\alpha\beta} = \frac{1}{2}(\partial_{\alpha}u_{\beta}+
\partial_{\beta}u_{\alpha}+
\partial_{\alpha} \! h \,\partial_{\beta} \! h)
\end{equation}
is the strain tensor. 

The harmonic approximation neglects  the non-linear
$ h^2 $ term, decoupling the bending and stretching modes.
Then the correlation functions for out-of-plane displacements,
$ H_0 (q) = \langle \vert h_{\bf q} \vert^2 \rangle_0 $, and for in-plane displacements,
$ D^{\alpha \beta}_{u,0} (q) = \langle u^*_{\alpha{\bf q}} u_{\beta{\bf q}} \rangle_0 $,
can be derived by Gaussian integration \cite{review,book}:
\begin{equation}
\label{H0q}
H_0 (q) = \frac{k_B T}{ \kappa q^4 }
\end{equation}
and:
\begin{equation}
\label{D0q}
D^{\alpha \beta}_{u,0} (q) =
\frac{P_{\alpha \beta} ({\bf q}) k_B T}{(\lambda + 2 \mu) q^{2}} +
\frac{( \delta_{\alpha \beta} - P_{\alpha \beta} ({\bf q}) )k_B T}{\mu q^{2}}
\end{equation}
 with $ P_{\alpha \beta} ({\bf q}) = q_{\alpha} q_{\beta}/q^2 $.
The average height fluctuation behaves as
$ \langle h^2 \rangle_0 = \sum_{\bf q} \langle \vert h_{\bf q} \vert^2 \rangle_0 \sim L^2 $, 
implying instability of a membrane as a flat phase.

Due to the large out-of-plane fluctuations, however,
the harmonic behavior is not valid for small $q$ and one has to keep
the $ h^2 $ term in eq. \ref{strtens}. Since $ H $ remains
quadratic in $ u $, these degrees of freedom can still be integrated out. 
This leads to a Hamiltonian in Fourier space which
is a function of $ h_{\bf q} $ only~\cite{book}:
\begin{equation}
\label{Ham2}
\tilde{H}  = \frac{1}{2} \sum_{\bf q} \kappa q^4 \vert h_{\bf q} \vert^2 +
\frac{Y}{8} \sum_{\bf q,k,k'} R ({\bf q,k,k'}) h_{\bf k} h_{\bf q-k} h_{\bf k'} h_{- {\bf q} - {\bf k'}}
\end{equation}
where $ Y $ is the 2D Young modulus and
$ R ({\bf q,k,k'}) = ({\bf q} \times {\bf k})({\bf q} \times {\bf k}')/q^4$.
The anharmonic, quartic term reduces the height fluctuations,
stabilizing the flat phase, effectively described by a
renormalized bending rigidity $ \kappa_R (q)\sim q^{-\eta} $ with positive $ \eta $.
Hence, the height correlation $ H(q) $ for $ q \rightarrow 0 $ has the same form as
$ H_0 (q) $ in eq. \ref{H0q}, but with $ \kappa $ replaced by
$ \kappa_R (q) $ \cite{Nelson1}. Likewise, $ D^{\alpha \beta}_u (q) $ can be described
by renormalized $ \lambda_R(q),\mu_R(q) \sim q^{\eta_u} $ in eq. \ref{D0q} with
$ \eta_u >0 $. From rotational invariance it follows that $ \eta $ and $ \eta_u $ should
satisfy the scaling relation $ \eta_u = 2 - 2 \eta $ \cite{aron}.

Within the self-consistent screening approximation (SCSA) \cite{Doussal},
the exponent was estimated as $ \eta \simeq 0.821 $ \cite{Doussal}; next-order
corrections reduce it slightly to $ \eta \simeq 0.789 $ \cite{Gazit}.
A renormalization group approach gives $ \eta=0.849 $ \cite{Kownacki} and
MC simulations for self-avoiding membranes
$ \eta \simeq 0.72 $ \cite{Bowick1}.
With $\eta > 0$, $ \langle h^2 \rangle \sim L^{2 - \eta} $, is much smaller
than $ \langle h^2 \rangle_0 \sim L^2 $ for large $L$, stabilizing
the flat phase.

Although it is {\it a priori} not obvious whether the
membrane theory applies to an atomic-layer-thick 2D crystal like graphene,
atomistic MC simulations confirm the scaling behavior of $ H(q) $  
with $ \eta \simeq 0.85 $\cite{Los1}. The scaling  of in-plane elastic moduli, 
however, has not yet been studied nor confirmed for graphene.
Contrary to $ \kappa_R $ which increases with increasing system size, making the membrane
more resistant against bending, $ \lambda_R $ and $ \mu_R $ decrease with
system size. Hence, if graphene follows the membrane theory, the 
in-plane elastic moduli vanish for large system sizes,
an unthinkable situation for 3D crystals!

For a 2D system,  the
2D bulk modulus $ B $, the uniaxial modulus $ C_{11} $
and $Y$ are related to $ \lambda $ and $ \mu $ as:
\begin{equation}
\label{Ym}
B = \lambda + \mu ~,~~ C_{11} = B + \mu ~~~
\mbox{and} ~~~ Y = \frac{ 4 B \mu }{ B + \mu }
\end{equation}
implying that $ B $, $ C_{11} $ and $ Y $ scale as $ \lambda $ and $ \mu $.
Another relevant quantity is the 2D Poisson ratio $ \nu $:
\begin{equation}
\label{Pr}
\nu = \frac{ B - \mu }{ B + \mu }
\end{equation}
The SCSA predicts a universal, negative Poisson ratio $ \nu = -1/3 $
for $ L \rightarrow \infty $ \cite{Doussal}, as  
later confirmed by MC simulation of self-avoiding
membranes~\cite{Bowick2}. For graphene, however, so far only
positive values were reported ($ \nu = 0.15 - 0.46 $)
\cite{Zakharchenko,Cadelano,Cao}.

\begin{figure}[htb]
\vspace*{0.00cm}
\includegraphics[width=7.cm,clip]{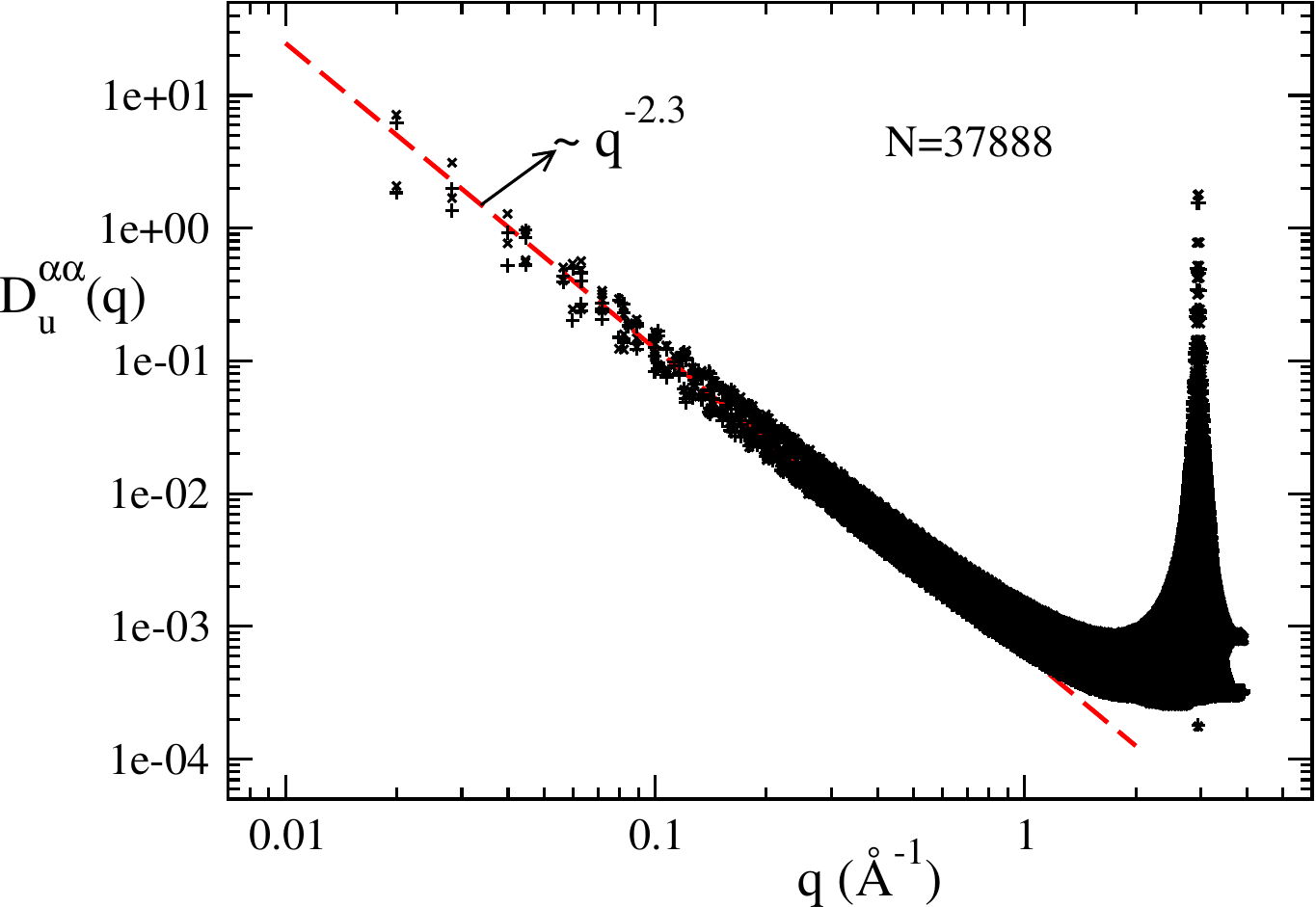}
\caption{ \label{Du}
Correlation functions $ D^{\alpha \alpha }_u ~(\alpha=x,y) $ for
in-plane displacements $ u_{ix} $ (x) and $ u_{iy} $ (+). 
The scaling exponent is consistent with 
$ D^{\alpha \alpha }_u \sim q^{-2 - \eta_u}$
with $ \eta_u = 2 - 2 \eta = 0.3 $, using $ \eta \simeq 0.85 $~\cite{Kownacki} (dashed line).}
\end{figure}

In Fig.\ref{Du} we show the in-plane correlation function $ D^{\alpha \alpha}_u (q)~(\alpha=x,y) $
calculated by $NPT$ MC simulations
at pressure $ P = 0 $ and $ T =$ 300 K with isotropic volume fluctuations for 
roughly square samples of $ N=37888 $ atoms using periodic boundary conditions.
Besides displacement moves we apply also collective wave 
moves for small $ q $  as in Ref.~\onlinecite{Los1}.
For the calculation of $ D^{\alpha\alpha}_u (q) = \langle \vert u^{\alpha}_q \vert^2 \rangle =
(1/N) \langle \vert \sum_i^N u_{i\alpha} \exp{(i {\bf qr}_{i,0} )} \vert^2 \rangle $
with $ \{r_{i,0}\} $ the ground state positions, the in-plane displacement
field was scaled as  $ u_{i\alpha} = s r_{i\alpha} - r_{i\alpha,0} $ where
$ s = \sqrt{A_0/A} $ scales
the area $ A $  at $T$=300 K to the ground state
area $ A_0 $ of a flat sample.
The behavior of $ D^{xx}_u (q) $ for small $ q $
is consistent with a power law with exponent
$ \eta_u \simeq 0.3 $, indicating that graphene follows the 
membrane theory also for in-plane correlations.
\begin{figure}[htb]
\vspace*{0.00cm}
\includegraphics[width=7.0cm,clip]{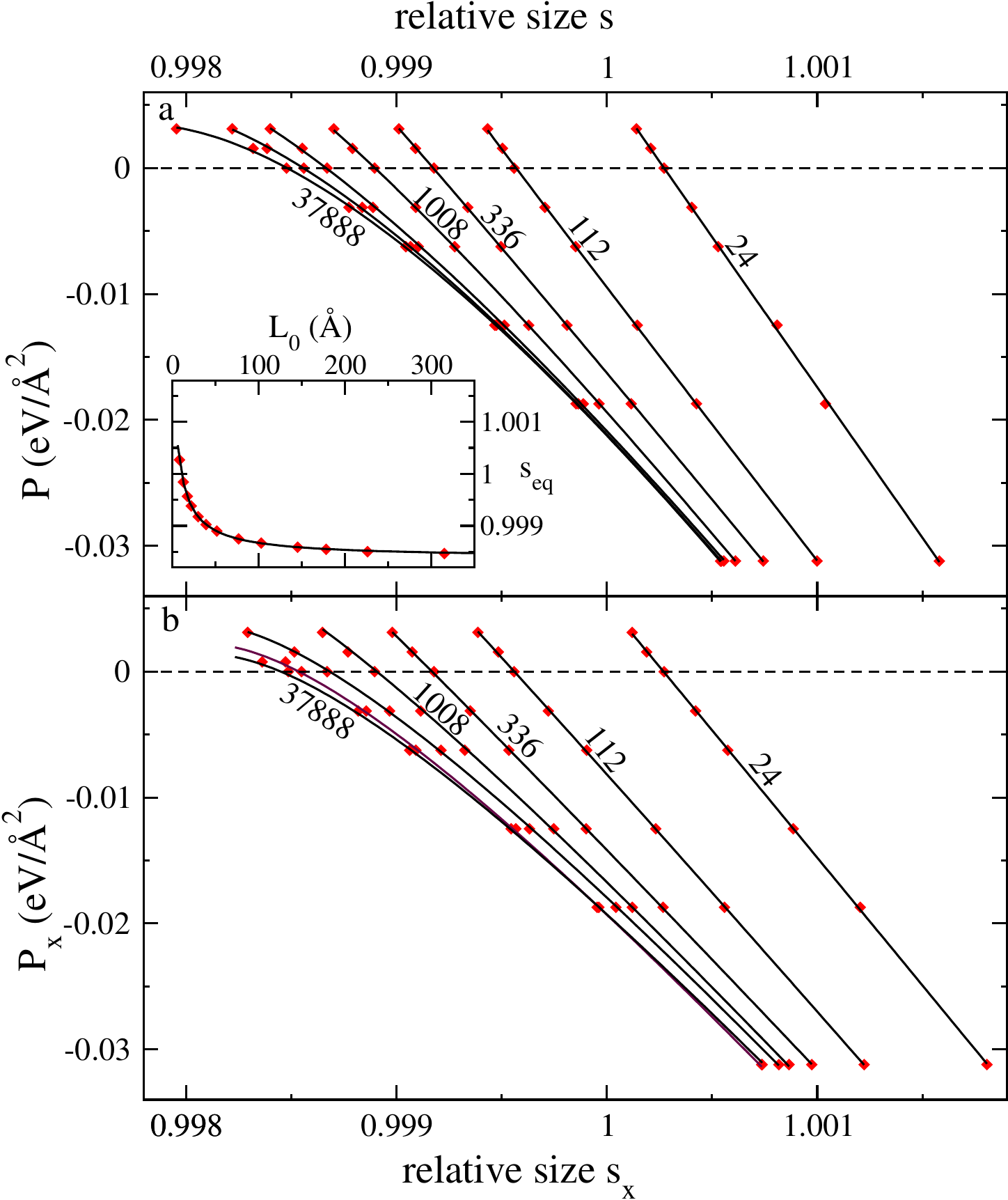}
\caption{ \label{EOS}
Pressure as a function of size in NPT simulations (symbols)
with (a) isotropic and (b) uniaxial size fluctuations
for approximately square systems with
$ N $ =24, 112, 336, 1008, 4032, 12096 and 37888 atoms.
The lines are best fits to eq. \ref{Ps}. The inset
gives the equilibrium sizes $ s_{eq} = s(P=0)$ (symbols) as a function
of $ L_0 $ and the fit (solid line) according to the expression in
Table \ref{Tab}.}
\end{figure}

The size and strain dependence
of the elastic properties can be computed simultaneously by
NPT MC simulations for different sizes with isotropic area fluctuations at several 
pressures $P$. The resulting
average area $ A $ gives the equation of state (EOS), $ A(P)$
and thus also $ P(A) $ from which $B$ can be calculated as:
\begin{eqnarray}
\label{bulkmod2D}
B = - A \frac{\partial P}{\partial A} = - \frac{s}{2} \frac{\partial P}{\partial s}
\end{eqnarray}
where $ s = L/L_0 $ is the relative linear system size 
with $ L_0 = \sqrt{N/\rho_0}$ the ground state system size, 
$ \rho_0 \simeq 0.3819$ \AA$^{-2} $ being the 2D ground state
atomic density of graphene. To obtain both $B$ 
and $C_{11}$, we also performed NPT simulations for uniaxial pressure $ P_{x} $, 
applying fluctuations  of $ L_x $ in the $x$-direction,
while keeping $ L_y $ fixed. Then $ C_{11} $ follows from:
\begin{eqnarray}
\label{C11}
C_{11} = - L_x \left . \frac{\partial P_{x}}{\partial L_x} \right|_P=
- s_x \left . \frac{\partial P_{x}}{\partial s_x} \right|_P \simeq
- s_x \left . \frac{\partial P_{x}}{\partial s_x} \right|_{s_{y}=s_{eq}}
\end{eqnarray}
where $ s_{\alpha} = L_{\alpha}/L_{\alpha,0} $, with $ L_{\alpha,0} ~(\alpha=x,y) $
the ground state dimensions, and where $ s_{eq} = s(P=0) $
is the equilibrium size obtained from the isotropic NPT simulations at $ P= 0 $.
The subscript ``{\it P}'' in eq. \ref{C11} indicates that $ L_y $ should be taken equal to $ s_y = s (P) $ resulting
from isotropic NPT simulations at pressure $P$ and that $ P_x $ should
be varied around $P$. However, since we verified that the dependence of
$ \partial P_x/\partial s_x $ on $ s_y $ is very
weak we adopted the last approximation in 
eq. \ref{C11}, which is exact for $ P=0 $.

The results are shown in Fig. \ref{EOS}.
The inset shows that the previously found negative thermal expansion~\cite{Zakharchenko}
is also size dependent, but tending to a constant for large $ L_0 $.
On the basis of Fig. \ref{EOS}a, with the slope $ \partial P/\partial s = 2B/s $
tending to a constant for large $s$, we propose the phenomenological relation for $B(s)$
\begin{eqnarray}
\label{Bs}
B(s) =  \frac{s \left( B_{eq}/s_{eq} + C D (s-s_{eq} ) \right) }{ 1 + D ( s - s_{eq} ) }
\end{eqnarray}
where $ B_{eq} $ is the equilibrium value at $ P = 0 $.
Substitution of eq. \ref{Bs} into eq. \ref{bulkmod2D} and integration yields
the EOS:
\begin{eqnarray}
\label{Ps}
P(s) = - \frac{2}{D}
\left( \frac{B_{eq}}{s_{eq}} - C \right) ln \left( 1 + D ( s - s_{eq} )\right)
- 2C ( s - s_{eq} ) \nonumber \\
\end{eqnarray}
Similarly, we write
\begin{eqnarray}
\label{C11s}
C_{11}(s_x) =
\frac{s_x \left( C_{11,eq}/s_{eq} + \tilde{C} \tilde{D} (s_x-s_{eq} ) \right) }
{ 1 + \tilde{D} ( s_x - s_{eq} ) }
\end{eqnarray}
which substituted in eq. \ref{C11} gives an equation for $ P_x (s_x) $
similar to eq. \ref{Ps} but with $s$, $ B_{eq} $, $ C $ and $ D $
replaced by $s_{eq}$, $ C_{11,eq}/2 $, $ \tilde{C}/2 $ and $ \tilde{D} $.
This form allows the excellent fits shown in Fig. \ref{EOS}, providing $ B_{eq} $ and
$ C_{11,eq} $ as a function of $ L_0 $.
\begin{figure}[htb]
\vspace*{0.00cm}
\includegraphics[width=8.75cm,clip]{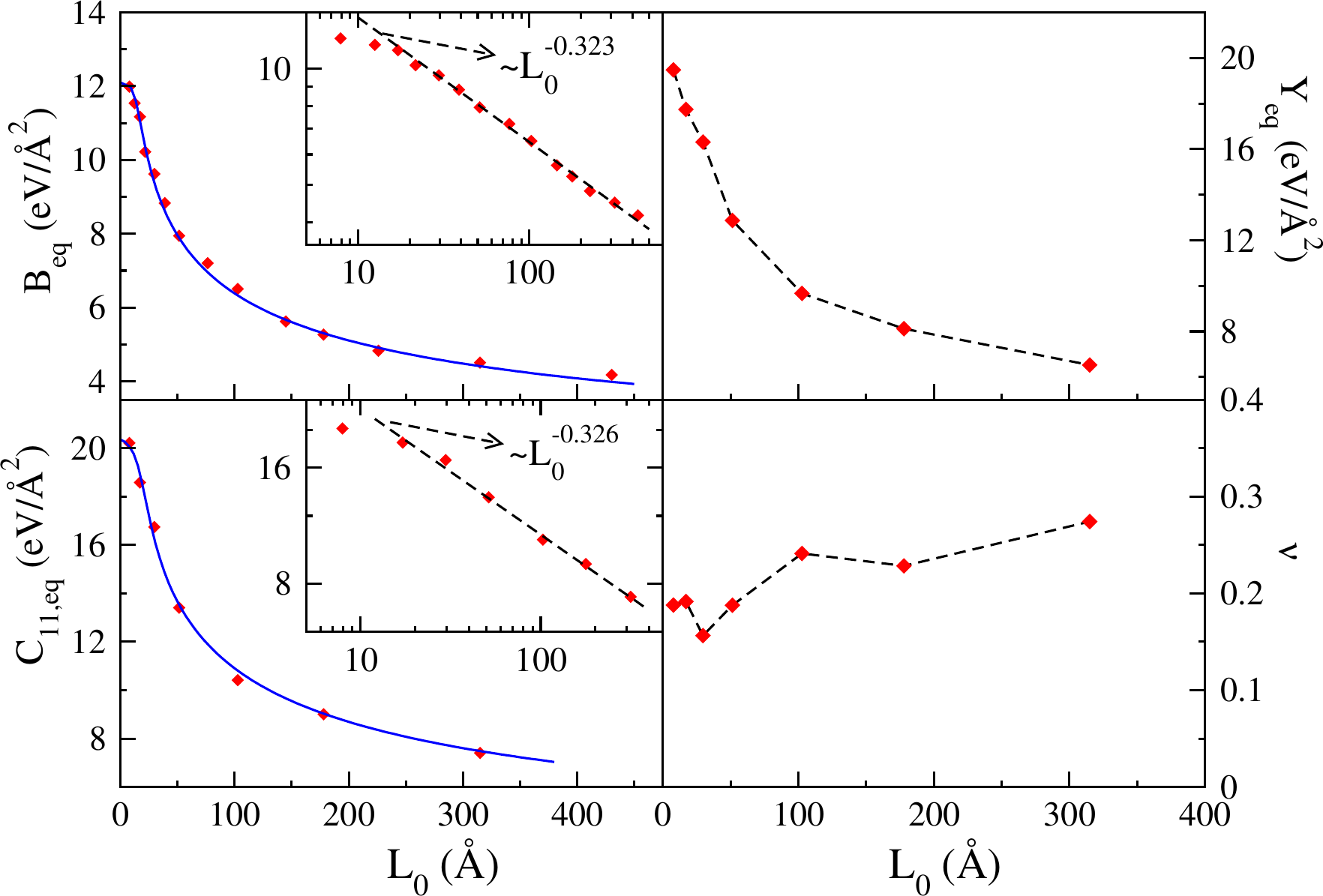}
\caption{ \label{BeqC11YmPr}
Equilibrium bulk modulus $ B_{eq} $, uniaxial modulus
$ C_{11,eq} $, Young modulus $ Y_{eq} $ and Poisson ratio $ \nu $
as a function of system size $L_0$. The insets in log-log scale
demonstrate the power law behavior. The solid
lines are fits according to the expressions in Table
\ref{Tab}. Dashed lines in the right panels are guides to the eye.}
\end{figure}
In the left panels of  Fig. \ref{BeqC11YmPr} (left panels) we show that both $B$ and $C_{11}$
vanish for large $ L_0 $, decreasing  as a power law  $ \sim L^{-\eta_u } $, with
$ \eta_u \simeq 0.325 $ (insets).
The right panels give the corresponding results for $Y$ and $\nu$ at
$ P=0 $, calculated using eqs. \ref{Ym} and \ref{Pr}.
Note that, according to LCBOPII, the in-plane elastic moduli of
graphene at $T=0$ K are $ B = 12.69 $ eV/\AA$^2$
and $ \mu = 9.26 $ eV/\AA$^2$, yielding $ Y = 21.41 $ eV/\AA$^2 = 343 $ N/m
and $ \nu=0.156 $, in agreement with {\it ab initio} data~\cite{Sanchez}
and with the small size limit  in Fig. \ref{BeqC11YmPr}
where $ Y \simeq 314 $ N/m. 
By simulations at 1 K for $N=24$ we verified that the remaining difference is due to temperature.

Interestingly, the power law decrease of $ B $, $ C_{11} $ and $ Y_{eq} $
as a function of $ L_0 $ sets in from $ L_0 \simeq 20$ \AA, a value
twice smaller than the Ginzburg critical value
$ L^* = 2 \pi \sqrt{16 \pi \kappa^2/(3 Y k_B T)} \simeq 40$ \AA~
(using $ \kappa \simeq 1.1 eV $~\cite{Fasolino1}) expected from membrane 
theory~\cite{Nelson1}. The Poisson ratio $ \nu $ for small sizes is close 
to its bare value and increases up to 0.275 for larger $ L_0 $, against
the SCSA prediction $ \nu = -1/3 $. Since
the scaling of $ B $ and $ \mu $ is consistent with the SCSA, it is very unlikely 
that $ \nu $ will reach the value -1/3 for $ L_0 \rightarrow \infty $, as the outcome 
of eq. \ref{Pr} only depends on the prefactors.
\begin{figure}[htb]
\vspace*{0.00cm}
\includegraphics[width=7.5cm,clip]{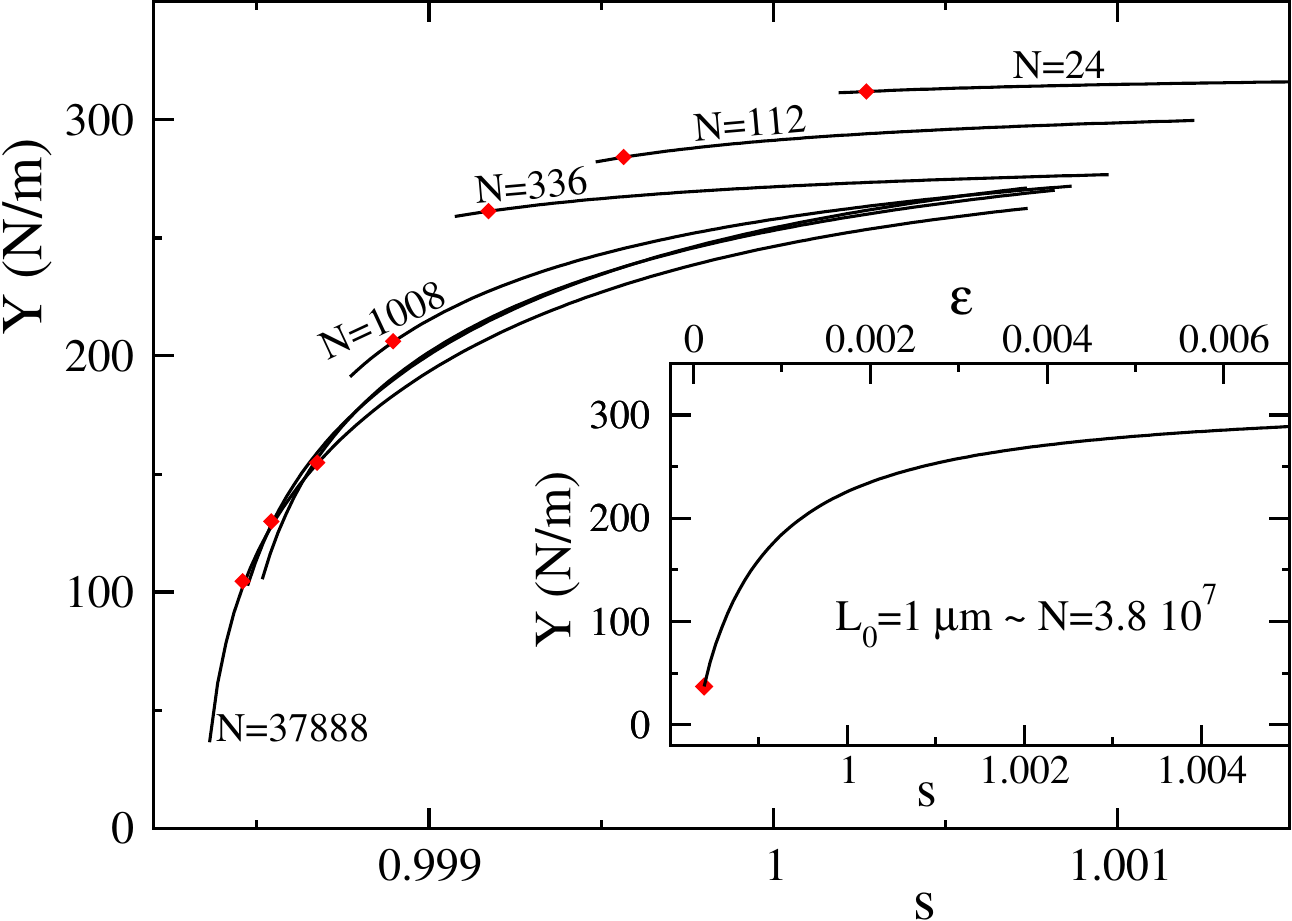}
\caption{ \label{Ys}
Young modulus $ Y(s) $ as a function of $ s $ for sample sizes as in Fig. 2 with symbols for
$ s=s_{eq} $.
The inset shows $ Y(s) $ for a size of
$ 1~ \mu m ~ (N \simeq 3.8 \times 10^{7}) $, calculated
using eqs. \ref{Bs} and \ref{C11s} and the expressions in
Table \ref{Tab}. The upper axis of the inset gives
the strain $ \epsilon= s - s_{eq} $.}
\end{figure}

We can also calculate $ Y $ as a function of tensile strain,
using eqs. \ref{Bs} and \ref{C11s} with the best fit parameters.
We should use the $ B(s) $ and $ C_{11}(s_x) $ at equal pressure by
solving $ P_x (s_x) = P(s) $ for $ s_x $ at given $s$. Due
to the approximation in eq. \ref{C11}, $ s_x \neq s $ unless $ s=s_{eq} $.
An approximation of $ s_x(s) $ is given in Table \ref{Tab}. 
The  $ Y(s) $ obtained from the data of Fig. \ref{EOS} are shown in Fig. \ref{Ys} for different  sizes. 
Symbols mark the
results at $ s=s_{eq} $. Notice the
strong increase of $ Y(s) $ for large sizes.
For  $ N = 37888 $ ($ L_0 \simeq 315 $ \AA),
$ Y $ increases from $ \sim $ 100 N/m at $ s_{eq} \simeq 0.9985 $
to 220 N/m at $ s=0.9995 $, i.e. more than a factor 2 for
a strain $ \epsilon = s-s_{eq} = 0.001 $ (0.1 \%)! This strong dependence is in 
full qualitative agreement with the recent experimental claims \cite{Polin1,Polin2}.
A quantitative comparison will be discussed below. 
\begin{table}
\begin{tabular}{l}
\hline \hline
$ s_{eq}=0.99838+\frac{4.295~10^{-3}}{1+0.1814 L_0^{0.94}}$, ~~
$ D=\frac{592.3 +1.2510^{-2} L_0^2 }{ 1 + 1.25 10^{-5} L_0^2 } $, \\
$ B_{eq} = \frac{12.1 - 5.69~10^{-3} L_0^2 + 28.6 (L_0/14.14)^4 L_0^{-0.325}}
{ 1.0 + (L_0/14.14)^4 } $, ~~ $ C = 12.1 $, \\
$ C_{11,eq} =
\frac{20.35 - 7.597~10^{-3} L_0^2 + 1.47~10^{-4} L_0^3 + 48.1 (L_0/31.62)^4 L_0^{-0.325}}
{ 1.0 + (L_0/31.62)^4 }, $ \\
$ \tilde{C}=20.35 $, ~
$ \tilde{D} = \frac{309.2 + 0.1597 L_0^2}{1+ 1.45 10^{-4} L_0^2} $,~
$ s_x(s) = 1.15 (s-s_{eq}) +s_{eq} $ \\
\hline \hline
\end{tabular}
\caption{Size dependent parameters for $ B(s) $ and $ C_{11} (s_x) $
according to eqs. \ref{Bs} and \ref{C11s} for 
$ L_0 $ in \AA.
$ B_{eq} $, $ C_{11,eq} $, $ C $ and $ \tilde{C} $ are in eV/\AA$^2$,
other quantities are dimensionless.}
\label{Tab}
\end{table}

Since $ B_{eq} $ and $ C_{11,eq} $, as well as $ s_{eq} $, $ C $, $ D $,
$ \tilde{C} $ and $ \tilde{D} $ turn out to depend smoothly
on $ L_0 $, we can approximate all parameters by the functions
of $ L_0 $ given in Table I. These expressions apply to
any size and give appropriate asymptotics with $ C $ ($ \tilde{C} $) equal to
 $ \partial P/\partial s $ ($ \partial P_x/\partial s_x $) for
the smallest system ($N$=24). 

The inset of Fig. \ref{Ys} shows the resulting $ Y(s) $ for a size $ L_0 = 1~ \mu m $ ($ \sim N=3.8 \times 10^7 $ atoms).
At zero strain (symbol)  $ Y $ is only 30 $ N/m $,  becoming almost a factor
10 larger at 0.5 \% tensile strain, where it approaches its asymptotic value.
Although the suppression of anharmonicity found here goes very fast as
a function of strain, it clearly deviates from membrane
theory within the SCSA, where a complete suppression of anharmonicity occurs for tensile strain 
of 0.01 \% \cite{Roldan}, two orders of 
magnitude lower than our atomistic simulations.

\begin{figure}[htb]
\vspace*{0.00cm}
\includegraphics[width=7.5cm,clip]{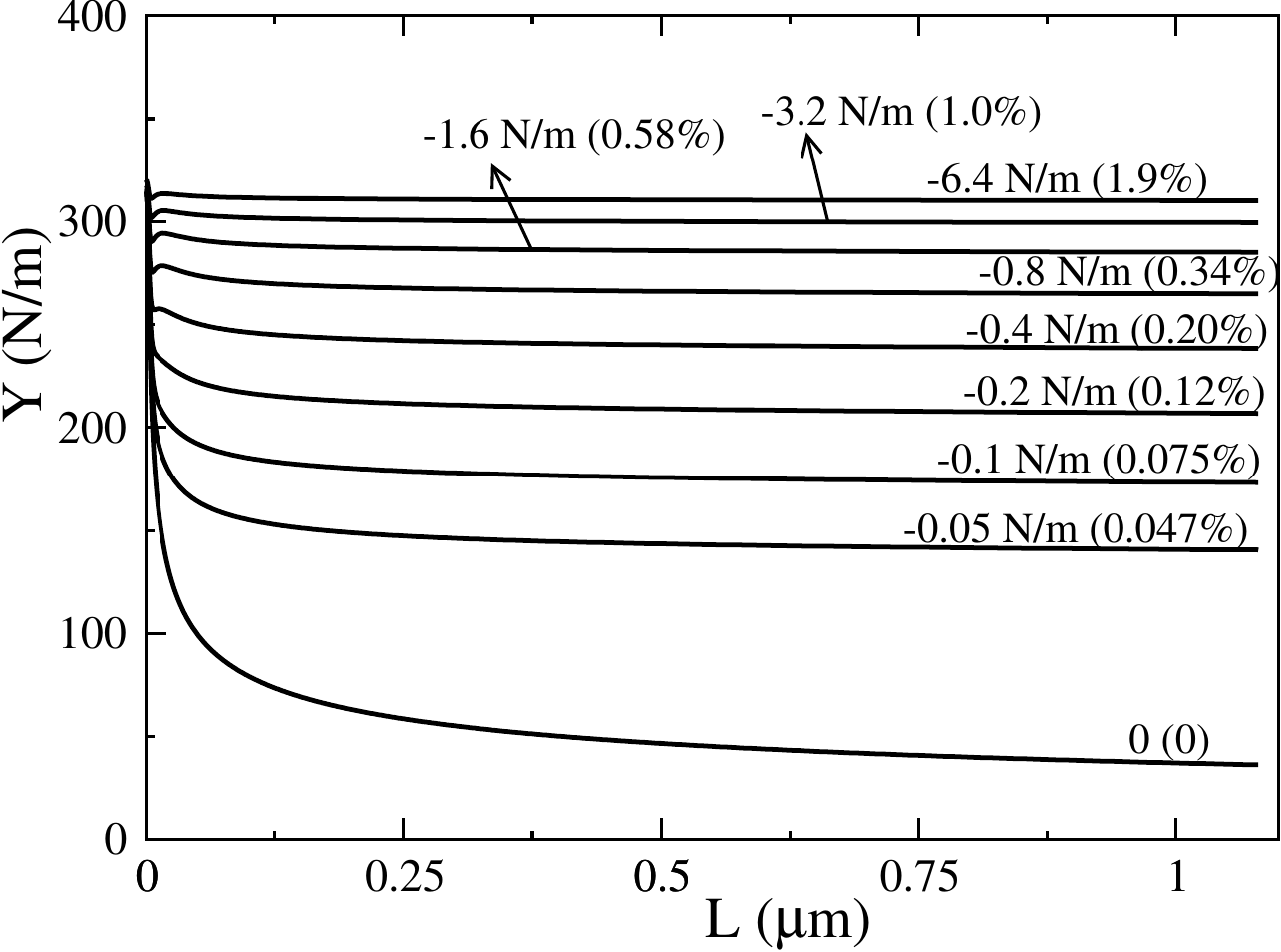}
\caption{ \label{YL0}
Young modulus $ Y $ as a function of  $ L_0 $ for the indicated 
values of the pressure $P$.
The value in brackets is the corresponding strain
$ \epsilon = s-s_{eq} $.}
\end{figure}
Finally, the size dependence of $Y$ with tensile strain
at  negative pressures is displayed in Fig. \ref{YL0}. Tensile stress
of 0.05 $N/m$, corresponding to $\sim$0.05 \% strain, suppresses the anharmonic
effects, and thus the power law decay, for
$ L_0 > 0.25~\mu m $. As a consequence, $ Y $ is a factor $ \sim 4 $ larger than
$ Y_{eq} $ for a system of 1 $ \mu m $. Subsequently, increasing the stress
by a factor 10 yields a strain of $ \sim $0.25 \% and $ Y \simeq 275 $ N/m.
This variation of $ Y $ with strain corresponds to recent
experimental data.\cite{Polin2} The factor 2 difference in both lower
and upper bound of $ Y $, however, with an experimental
upper value $ Y$ =700 N/m, remains unexplained and requires further 
investigations. We note that our values of 
the bare elastic moduli agree with the experimental phonon spectrum~\cite{phonon1,phonon2} of graphite, where
anharmonic effects are suppressed by interlayer interactions.  

In conclusion, we have shown by atomistic simulations that the in-plane
elastic moduli of graphene vanish with size as $ L_0^{-\eta_u} $ with $ \eta_u \simeq 0.325 $, 
confirming that graphene follows membrane theory within the
SCSA in this respect. The critical exponent $ \eta_u $, 
together with the independent  estimate of $ \eta \simeq 0.85 $~\cite{Los1},  
supports the scaling relation $ \eta_u = 2 - 2 \eta $.
In contrast, our results do not support the SCSA
prediction $ \nu = -1/3 $ for $ L \rightarrow \infty $. We remark that $ \nu = -1/3 $ in eqs. \ref{Ym} and \ref{Pr}
 leads to $ B_R = -\lambda_R $ and $ \lambda_R = 2B_R -C_{11} $, implying 
 that $ \lambda_R $ should be negative for stability while in
Fig.~\ref{BeqC11YmPr} $ 2B_R - C_{11,R} $ remains positive
for any $ L_0 $. We also find that suppression of anharmonicity requires a tensile strain 
10-50 times larger than predicted by SCSA.

\begin{acknowledgments}
This research has received
funding from the European Union Seventh Framework Programme under grant agreement
n°604391 Graphene Flagship.
\end{acknowledgments}

\end{document}